# Investigating student perceptions of creativity and generative ai in computational physics

Pachi Her and Patti Hamerski

*Department of Physics, Oregon State University, 1500 SW Jefferson Way, Corvallis, OR 97333*

Generative Artificial Intelligence (gen-AI) is rapidly becoming more integrated into today's classrooms in all ranges of education. In higher education, Gen-AI is often seen as a resource for students, aiding them in drafting outlines, solving simple mathematical problems, or even decoding or constructing code. In this paper, we analyze essay-based interviews (N=6) from an upper-division computational physics course, in which physics majors addressed their views and attitudes towards Gen-AI and how it affects their learning. We analyzed the concepts of creativity and gen-AI using the Four C Model, a framework encompassing four types of creativity. Our analysis of the data involved coding and characterizing students' definitions of creativity and generative AI. Our findings revealed two main observations: first, students conceptualized their creativity primarily within mini-c and little-c; second, students perceived gen-AI as a resource and learning tool but expressed skepticism regarding its accuracy and creativity.

## I. INTRODUCTION

Generative artificial intelligence (gen-AI) are AI systems capable of generating various types of content, including text, images, audio, and video, based on their training data and user inputs [1]. These systems respond to prompts which can be either unimodal (single type of input) or multimodal (multiple types of input). An example is ChatGPT, short for "Chat Generative Pre-trained Transformer," developed by OpenAI. Notably, ChatGPT simulates human-like conversations in real-time. Several implementations of gen-AI include Copilot, Gemini, Jasper, and many others.

Given its versatility and recency, gen-AI is increasingly prevalent in today's education landscape, spanning various grade levels from K-12 to college. It is used in curriculum development by teachers and consulted for homework help by students [2]. While research on gen-AI in education is still relatively limited, it is gaining traction across various fields [3, 4]. Similarly, research within the Physics Education Research (PER) community on gen-AI is expanding rapidly, with a focus on assessing the quality and accuracy of AI responses in physics tasks [5–9]. For instance, Sirnoorkar et al. examined and compared responses from students and ChatGPT on specific physics problems to investigate their sense-making and mechanistic reasoning processes [10]. Polverini and Gregorcic assessed ChatGPT's performance in interpreting kinematic graphs, highlighting both its limitations and capabilities [11]. Much of the work done in PER has focused on the high school to lower-division physics level. We could not find any research that has looked into the role of gen-AI in computational physics courses, nor upper-division undergraduate physics contexts.

Our study is motivated by two key factors. Despite limited exploration at the intersection of gen-AI and physics, particularly within computational physics, the inherent problem-solving and modeling nature of computational physics aligns well with gen-AI capabilities. The integration of gen-AI as an advanced search tool, akin to common practices like googling and debugging in computing, presents potentially useful applications for gen-AI. With computational physics gaining traction in both research and education, there is a growing need for innovative tools and methodologies to enhance teaching and learning in computational courses [12].

Second, as gen-AI increasingly permeates today's education, its relationship with creativity development becomes more pronounced. A growing body of research has delved into this relationship across various contexts [13]. Marrone et al. used the Four C model of creativity to investigate middle school students' perceptions of AI and creativity. Their work serves as an inspiration for our study as they called for further exploration on AI and the Four C model in other education contexts, especially in the development of mini-c and little-c creativity, where "the creative output is not as crucial as the self-discovery that occurs through the creative process." [14]. Thus, the research question of our study is: *How are students' conceptual framings of their creativity connected to their perceptions of gen-AI in a computational physics education context?*

## II. THEORETICAL FRAMEWORK

The Four C Model by Kaufman and Beghetto serves as our theoretical framework and aids with characterizing our students' perceptions of creativity [15]. The framework emphasizes that creativity is a developmental quality rather than innate. This framework was constructed to help conceptualize and classify the different levels of creative expression. As the name of the framework implies, there are four types or levels of creativity: mini-c, little-c, Pro-C, and Big-C. Our use of this model focuses mainly on mini-c and little-c, the types of creativity most closely associated with students' engagement with the creative process [14]. This model demonstrated its relevance for understanding gen-AI applications as mentioned in the previous section by Marrone et al..

Kaufman and Beghetto defined mini-c as the "novel and personally meaningful interpretation of experiences, actions, and events." This type of creativity centers the individual, who finds meaning and interest in their creative product. The creative product typically does not get shared or exposed to the social surroundings. Kaufman and Beghetto introduced this type of creativity to "encompass the creativity inherent in the learning process." The next type of creativity is little-c which follows the standard definition of creativity from Plucker et al: "the interaction among aptitude, process, and environment by which an individual or group produces a perceptible product that is both novel and useful as defined within a social context." [16]. A key element that differentiates little-c from mini-c is the social factor.

In Pro-C and Big-C, the creator is a domain expert, which lies in contrast to the students in this study who are early in their physics career. For this reason, the nuances between Pro-C and Big-C are shown in Table 1 but not highlighted here because we observed very few excerpts from the data that reflected Pro and Big-C levels, which we will address in the last section of this paper.

## III. METHODS

We took an interpretive approach to data generation and analysis because we wished to explore how students viewed the role of gen-AI in their creative pursuits. This approach led to the methods below, which centered students' interpretations of gen-AI [17]. Taking a qualitative approach and focusing on a single context were intentional choices for this work because qualitative research on single cases can capture the narrative of students' attitudes, understanding, and use of gen-AI in their creative works with the detail needed for an understudied area such as this [18]. This paper explores how students in a computational physics course perceive gen-AI and the extent to which they have used gen-AI in their creative processes.

## A. Data Generation

The data we collected took place at Oregon State University in a series of upper-division computational physics courses called PH 36X. This course sequence spans three-terms with the majority of enrollment being physics majors in their junior year. The course sequence aims to teach the fundamentals of programming in Python and how to utilize computing as a tool for doing and learning about physics phenomena. Students encounter a variety of physics topics with tasks intentionally designed as open-ended, with the goal of fostering experimentation and discovery of diverse solutions. During the time of the study, the participants were enrolled in the second term of the course, PH 366. Six students consented to partake in this study and were compensated with a $20 Amazon gift card. These students had varying levels of computational experience. Of the six participants, one had taken a computer science course (data structures) prior to this course sequence, one had informal programming training, and all six had taken or waived the first course (PH 365) of this course sequence, a one credit course that covers introductory programming concepts in python and their applications to physics problems.

The data we collected were narrative essays, which consisted of three essay prompts that addressed topics on gen-AI/ChatGPT, creativity, and intuition. Our intention of conducting narrative essays was inspired by the work from Lo-Presto and Drake, where they explained how collecting mathematics stories or narratives from preservice and in-service math teachers can provide helpful insight and understanding in their relationships to the field of mathematics, aiding with curriculum design and instruction [19]. Our aim with using narrative essays is similar such that we wanted to capture the students' perspectives in their words. Although conducting interviews could potentially evoke the same objective, narrative essays can be advantageous by allowing students to reflect and also reevaluate what they are writing about.

The first essay prompt asked students about their views and opinions on gen-AI. The second prompt asked students to define creativity (and intuition) in their own words along with sharing any examples of when they were creative in [any of] their physics course(s). In the third prompt, we brought the two concepts of gen-AI and creativity together and asked students to describe the relationship between them and explain whether gen-AI affects their creativity. The prompts were drafted and fleshed out after the authors wrote various drafts of the questions based on literature review and classroom experience, and solicited discussion-based feedback from the broader physics education research group at Oregon State University.

The participants were asked to meet with the first author individually via Zoom or in-person to participate in the narra-tive essays. Participants were instructed before writing to respond to each prompt in the order presented to them and were encouraged to express their genuine feelings and share any experiences they were comfortable sharing for each prompt. The order of the prompts was placed in the specific sequence presented earlier to have students think about gen-AI and creativity separately. The purpose of this was to not prime students to connect the two topics and let them discuss the two without forcing a connection before reaching the last prompt. Each participant completed the essay prompts within an hour.

## B. Data Analysis

The Four C Model was used to code the responses. However, we expanded the definitions to accurately reflect and characterize the student responses with the incorporation of gen-AI, as seen in Table 1. The analysis process involved iterative stages of reading, interpreting, and coding the written responses. Initially, the unrevised definitions of the Four Cs were applied in the first round of analysis. An illustration of this initial analysis process can be seen when we analyzed Bruce's response to prompt 2, to which he responded:

> "Creativity comes from when you're completing a task, and *you find a way to do it without following a step-by-step formula*. Maybe this means *you found an interesting string of code to solve a problem where you weren't explicitly trained to solve that specific problem*. It could also mean that **you added your own personal flair to something**, like art or writing. Maybe it means you've been given a complicated problem that you've never seen before and *you have to piece together all the skills* you've learned in non-obvious ways." (emphasis added by the authors)

Applying the original definitions of the Four C model, we categorized Bruce's response as closely aligned with the concept of mini-c. The **bolded phrase** represents our initial coding of mini-c. Bruce made the comment of adding "your own personal flair to something," resembling the original definition of mini-c as "the novel and personally meaningful interpretation of experiences, actions, and events," wherein individuals incorporate their personal interpretation or style.

The following stages of analysis involved iteratively expanding the Four C's definitions and recoding the written responses. The *italicized phrases* emphasize the expanded mini-c elements in the revised definitions to incorporate the notion of learning as an inherent component of the creativity process, particularly at the mini-c level. Finally, we incorporated gen-AI into the definition when we reviewed prompts 1 and 3 which asked about perceptions of gen-AI. Bruce responded to prompt 1 with the following:

> "Looking at where generative AI is today, I think that it has a lot of value to be used as a tool...It

Table I. The expanded definitions of the 4C Model. The expanded version also includes how gen-AI fits into each type of creativity.

| 4C Model (gen-AI/ChatGPT) | Definition | Example |
|---|---|---|
| mini-c | The person goes through a series of learning experiences, **where gen-AI is used as a learning tool or resource of information and knowledge** and creates a product that is not necessarily novel or innovative (to others). The product is creative and meaningful to the individual and does not receive feedback from others. | "Writing a question into google might pop up a bunch of unrelated things or advertisements, but *generative AI would be more likely to give a helpful response, and you can also ask it to further elaborate. It can also be used to help with creative processes, like brainstorming or formatting.*" - Bruce, Prompt 1 |
| little-c | The person presents their creative product to their social surroundings (classmates, teachers, friends, etc.). **Gen-AI is used to simulate this social surrounding, partaking in the "social interaction" with the individual by generating feedback towards the individual's prompt.** Assessment of the creative product moves beyond the individual's intrapersonal interpretation and is open to the social surroundings to be evaluated. Direct or indirect feedback can potentially be generated by the surroundings. | "AI is especially good at *finding the errors in a chunk of code, then explaining exactly why the code is malfunctioning.*" - Summer, Prompt 1 |
| Pro-C | The person is at the professional level of creativity. The creative product is recognized by a larger social group, such as the field of study and impacts the field. **Gen-AI, recognized by the person, is viewed as an equal regarding skillset and its ability to function similarly or possibly more advanced than the individual.** | "A common thought among computer scientists right now is job security due to *AI being able to code much faster and easier.*" - Mateo, Prompt 1 |
| BIG-C | The creative product has a lasting impact to the field. Similarly, **gen-AI is recognized by its influence on the field, where its contributions impact the field in an everlasting manner.** | "*If we don't want AI to take over, if we want our existence as humans to mean something,* we need to be able to do things that Ai can't (like think critically, solve complex problems, etc.) and when students get used to using thinks like Chat GPT, it's getting rid of that "usefulness". - Ali, Prompt1 |

can also be used to help with the creative process, like brainstorming or formatting."

Bruce mentions the learning process that we had included in the definition for mini-c (brainstorming, exploration and gathering information, etc.) realizing the role gen-AI plays in mini-c. Table 1 was the final version of our codebook for the analysis presented in this paper.

## IV. RESULTS

We coded all three essay prompts for each participant. Based on the data, two main observations were made.

1. **Students framed their creativity around mini-c and little-c.** Several participants expressed the idea that creativity stemmed from the individual's prior experience, reflecting mini-c's emphasis on the individual as well as the learning involved in the creative process. The following is an example from the data that reflected mini-c.

> "Looking further into the word it almost seems like creativity is the ability or being able to create something. Looking even further, it could also be being able to create something new...One could say we are not necessarily creating something new, but we are creating a solution that is new to us." – Robin, Prompt 2

Robin defined creativity as the process of "being able to create something" and further commented that "One could say we are not necessarily creating something new, but we are creating a solution that is new to us," demonstrating the individual nature of mini-c. Little-c, akin to to mini-c but with a social component, extends beyond individual perspective to include the social context where the creative product is shared with the individual's social circle. Here, the creative product is shared with the individual's social circle. An example demonstrating little-c follows.

> "I would define creativity as the ability to think outside the box or find new/innovative ways to solve a problem...Sometimes in PH [blinded] one of my classmates sitting near me may have gotten the same answer or end result by organizing their code differently or using different functions to perform the same tasks, which is why it's important to collaborate and combine our creativity with others." – Summer, Prompt 2

In this excerpt, Summer initially defined creativity as a product that is novel and innovative. She then recalled instances in her computational physics course that aligned with little-c, emphasizing collaboration and combining creativity with classmates. She noted that her classmates sometimes arrived at the same solution despite writing different codes from hers. This illustrates little-c's social aspect where creative products, such as the codes that Summer described in her response, are being shared to the social environment, the classmates.

The remaining participants echoed Robin and Summer's definition of creativity. Luke, Mateo, and Ali for instance, defined creativity similarly to Summer, using words such as "novelty" or "thinking outside the box," highlighting the importance of the product being new and original. We also observed their responses mirroring little-c's characteristic of social recognition. Conversely, Bruce's definition closely resembled to Robin's response, highlighting mini-c elements as we saw in the previous section.

2. **Students possessed distinct views on human creativity and gen-AI creativity: while gen-AI is viewed as a supportive resource during mini-c, some students saw it as unreliable for little-c creativity.** The second observation addresses the attitudes towards human creativity and connections between gen-AI and creativity, examined in prompts 1 and 3. Responses to these prompts predominantly fell into the mini-c category, viewing gen-AI primarily as a tool for independent learning (e.g., brainstorming, information gathering, and outlining ideas). Despite students utilizing gen-AI as mini-c, we observed a theme in their responses expressing doubt and skepticism regarding the outputs that gen-AI produces. We present several examples illustrating these observations.

> "AI can be used to generate ideas or help brainstorm, but we should always consider what it says with a grain of salt (or a few), especially since this is still a relatively new technology & more keep coming at. Sometimes AI will give a response that's against what my intuition would tell me is true, so we definitely need to be wary of twisting everything it generates." – Summer, Prompt 3

Here, Summer illustrates the use of gen-AI as a learning tool, aiding her in generating and conjuring ideas. However, she immediately expresses caution regarding the quality of gen-AI's responses, writing, "we should always consider what it says with a grain of salt (or a few)". We observed that this cautious approach stemmed from her past experiences where gen-AI conflicted with her intuition causing her wariness of gen-AI's outputs. Similarly, we see caution from Luke who also utilized gen-AI as a learning assistant but voiced skepticism and disappointment over its usage.

> "The times I have used generative AI for creative inspiration have usually ended in disappointment. I once had an essay to write and, to experiment, asked ChatGPT to perform various aspects of the process (brainstorm, outline, draft, revise). The first three steps are more "generative, and its outputs were unoriginal and uninspiring. On the other hand, it was useful for the less creative task of clarifying sentence structure." – Luke, Prompt 3

We encounter another mini-c perspective on gen-AI from Luke. Luke utilized gen-AI for various learning tasks, including brainstorming, outlining, drafting, and revising. However, echoing the recurring theme from the previous excerpt, Luke expressed disappointment in gen-AI, noting that its generated products were "unoriginal and uninspiring." This sentiment mirrored the distrust and skepticism observed in other participants like Summer. Additionally, Luke's response also reflected little-c when he mentioned "revising" as one of the ways of using gen-AI, which must involve him sharing his work with ChatGPT. His final remark reflects the notion that while gen-AI may fall short in fostering creativity, it can still provide utility in certain aspects of the learning process.

We found that the majority of participants utilized gen-AI in mini-c ways, but we did observe a few instances of little-c. In Summer's response to prompt 1, she had shared that she had used gen-AI to review her code and praised it for finding errors but critiqued its ability in providing workable code. This demonstrates gen-AI's feedback is akin to sharing one's product with their social surroundings.

## V. DISCUSSION & CONCLUSION

Our results presented how our participants conceptualize creativity and how they feel about the use of generative AI in a computational physics classroom. We observed a pattern of skepticism amongst the responses, revealing to us that our participants are using gen-AI in mini-c ways but have learned that gen-AI cannot always be trusted, warranting being critical of its outputs. This demonstrates that our participants are recognizing the important nuances of how to use gen-AI effectively. However, one student, Robin, felt contrastingly and said that gen-AI can be creative and has the ability to create new content from prior 'experiences'. We observed that Robin humanized AI, communicating the idea that AI has prior experiences like humans. This observation poses the question of whether students humanizing gen-AI will lead to more acceptance of the tool and lead to having a more positive attitude towards it. Additionally, it questions whether it could lead to undue conflation of AI and human contributions to creative processes, potentially revealing insights about human creativity. This narrative of humanizing AI should be studied further to provide deeper insight on the roles of how our views on gen-AI can impact our creativity.

With only six participants, we recognized this study has its limitations as the results we have observed may not be generalizable to larger populations; the few remarks of Pro-C and Big-C could potentially be due to the small data collection. However, this encourages for further work in collecting more data on students' perspectives to be able to reach more broader claims. As our work stands now in the research presented here, this narrow focus allowed us to examine how our participants conceptualized mini-c and little-c creativity in detailed and nuanced ways. Additionally, investigating how students are prompting gen-AI may be another potential research to observe the interplay between a student's efficacy of the tool to their perspective of gen-AI and creativity.

This study presents our initial exploration into how students perceive and utilize creativity and gen-AI's capability within a computational physics course. The demonstrated interaction between creativity and utilization of gen-AI has implications for education as gen-AI continues to develop and become more prevalent. Our aim with this paper is to catalyze further studies and inquiries on the effects and dynamics of gen-AI and creativity in computational physics. This work builds on Marrone et al's research to investigate the connection between creativity and gen-AI among physics majors [14]. We examined how gen-AI could either bolster or constrain human creativity, and focus particularly on its implications within the realm of computational physics.

## ACKNOWLEDGMENTS

The authors wish to express their appreciation to the PER team at Oregon State University for their feedback and support.